\documentstyle[preprint,prd,aps,epsfig,12pt]{revtex}

\begin{document}
\preprint{}
\draft
\title{Transverse $\tau$ polarization in $B_c \to \tau
\bar \nu_{\tau} \gamma $ decay}
\author{Chao-Hsi Chang$^{1,2}$, A.K. Giri$^3$, Rukmani Mohanta$^4$ and Guo-Li Wang$^{1,5}$}
\address{$^1$ CCAST (World Laboratory), P.O.Box 8730, Beijing 100080, China.\\
$^2$ Institute of Theoretical
Physics, Chinese Academy of Sciences, P.O. Box 2735, Beijing
100080, China.\\
$^3$ Department of Physics, Panjab University,
Chandigarh-160014, India.\\
$^4$ School of Physics, University of
Hyderabad, Gachibowli, Hyderabad-500046, India.\\
$^5$ Department
of Physics, Fujian Normal University, Fuzhou 350007, China.\\}
\maketitle
\begin{abstract}
The possible transverse polarization of the $\tau$-lepton, which is
$T$-odd and can be used to measure $CP$-violation, is
estimated precisely in the radiative decay mode
$B_c \to \tau \bar \nu_\tau \gamma$ with possible parameters
for the multi Higgs doublet model and the R-parity violating
supersymmetric model. We find that the up-bound of the transverse
polarization with possible parameters in the models is at very
different levels i.e. it can be $\leq 15\sim 23\%$ for the former but $\leq
0.1 \%$ for the latter, and it is accessible experimentally soon.
\end{abstract}


\section{Introduction}
The $B_c$ meson has attracted special attention as it contains two heavy
quarks and its decay channels are very
rich compared to those of $B_q$ $(q=u, d, s)$ mesons. Its
production \cite{ref1}, spectroscopy \cite{ref2} and
different decay channels \cite{ref3} are widely discussed
in the current literature. Recently, it has been
observed in 1.8 TeV $p\bar p$ collisions using
the CDF detector on the Fermilab Tevatron and the values: mass
$M_{B_c} = 6.40\pm 0.39 \pm 0.13$ GeV and lifetime
$\tau_{B_c}=0.46 ^{+0.18}_{-0.16} \pm 0.03$ ps have
been obtained \cite{ref4}. Further
detailed experimental studies will be performed on
Tevatron Run II and CERN LHC, which is the motivation for
extensive theoretical studies of this system. In particular,
on LHC with luminosity ${\cal L} = 10^{34} cm^{-2}
s^{-1}$ and $\sqrt s $ = 14 TeV, the number of $B_c^{\pm}$ events
is expected to be about $10^8 \sim 10^{10}$ (or
even more)\cite{ref5}, so some interesting rare decays could be
studied experimentally in the foreseen future.

The comparatively `long' lifetime of the $B_c$ meson is related to
the fact that it can decay only weakly, thus providing the rather
unique opportunity of investigating weak decays in a heavy
quarkonium-like system. Its weak decay may be due
to three `components':

i. The $\bar b$ quark decays with the spectator $c$ quark
(e.g. in the decay $B_c \to J/\psi l \bar {\nu_l}$ etc);

ii. The $c$ quark decays with the spectator $\bar b$ quark
(e.g. in the decay $B_c \to B_s l \bar {\nu_l}$ etc);

iii. The annihilation of $\bar b$ and $c$ (e.g. the decay $B_c \to
l^+ {\nu_l}, (l= e, \mu, \tau)$ etc).

\noindent
Sometimes there is interference of two or three of
the above processes in certain nonleptonic decays.

As far as the annihilation processes are concerned, the pure
leptonic decay to $e^+$ or $\mu^+$, due to helicity
suppression, represents a minor fraction of the $B_c$ full width
but the decay to $\tau$ does not. Whereas all the pure leptonic
decays are the simplest and can be used in determining the
decay constant $f_{B_c}$. Thus in fact it is very difficult to use
the $B_c \to l^+ \nu_l, (l = e, \mu)$ decay modes to determine
$f_{B_c}$, but it is possible
in principle to use the channel $B_c \to \tau^+ \nu_{\tau}$ if we
 have good efficiency in the detection of
$\tau$-leptons.

The helicity suppression can be avoided, if there is a third
particle in the final state. In particular, `adding a photon' to
the corresponding pure leptonic decay does not appreciably change the fact
that the decay ratio is proportional to the decay constant
$f_{B_c}$, but the decay is changed to a `radiative one'. The radiative
leptonic decay modes $B_c \to l^+ \nu_l \gamma$ have already  been
studied using the quark models \cite{ref6}, light cone QCD
\cite{ref7} and effective field theory \cite{ref8}.

In this paper we intend to study the $B_c \to \tau^+ \nu_{\tau} \gamma$
decay mode. This mode is of particular interest because it can provide
quite a sensitive test of certain theories of $T$ or $CP$ violation. Since the
transverse $\tau$ polarization is proportional to $(\vec p_\tau \times \vec k)\cdot
\vec s_\tau $ where $\vec p_\tau $ and $\vec k$ are the momenta of the $\tau$ and
photon particles, respectively, and $\vec s_\tau $ is the spin of $\tau $, it is an
odd quantity under time reversal. The standard model (SM) has only negligible
contribution to the polarization in the decay mode\footnote{In SM, at tree level
the polarization is null.}, therefore measurement of the transverse $\tau$
polarization will reveal to us possible new sources of $T$ or $CP$ violation
beyond SM (assuming $CPT$ symmetry). Namely, many models (beyond SM) can contain
certain new physical phases which induce $CP$ violation, and measurement of the
transverse $\tau$ polarization in this decay mode is possible ($\tau$'s lifetime
in the order of $0.3 ps$ and weak interaction decay). If a nonzero value
of the polarization is observed, it will be a clear indication of the existence of
new $CP$ phases beyond SM. Here, as two examples to see the possibility or advantages in
observing certain new sources of $CP$ violation, we consider the multi Higgs
doublet model (MHDM)\footnote{Sometimes people call the two Higgs doublet model as
a MHDM too, but here we do not. Namely here we precisely mean that there are more than
two Higgs doublets in the model.} and R-parity violation models of the supersymmetric
extensions of SM, because in these models new $CP$-violation phases are
rich\cite{ref9,ref12} and the transverse polarization of $\tau$-leptons in the
$B_c \to \tau^+ \nu_\tau \gamma$ decay mode may be at tree level induced by the
$CP$ phases. Therefore throughout the paper, we restrict ourselves to estimate the
effects in the two models just at tree level.

The paper is organized as follows. In section II we obtain the transition amplitude
for the $B_c \to \tau \bar \nu_\tau \gamma$ decay using the nonrelativistic quark model.
The standard model contribution to the amplitude corresponding to $W$-exchange, the MHDM
additional contribution (charged Higgs exchange) and the RPV contribution (slepton
exchange) are explicitely given in subsections A, B and C, respectively. The formalism for
$\tau$ polarization is presented in section III. Section IV contains numerical results and
discussions.

\section{Transition amplitude for the $B_c \to \tau^+ \nu_{\tau}
\gamma$ decay process}

According to the Feynman rules for the SM, MHDM and
RPV-SUSY models, the relevant Feynman diagrams for the decay $B_c \to
\tau^+ \nu_{\tau} \gamma$ are obtained to be just those with a photon
attached in turn to each of the charged fermion lines of the
diagrams  for the pure leptonic decay $B_c \to \tau^+
\nu_\tau$, as shown in figures 1, 2, 3-a, b, c, d, respectively. The matrix element
can thus be written down immediately. Based on the experimental facts,
in the models charged gauge bosons$W^\pm$ and the concerned Higgs particles
or sleptons are heavy compared to $B_c$, so we do not take
into account the contributions arising from the diagrams where the
photon is attached to the mediating charged bosons (figures 1, 2, 3-d). This
is because they are suppressed at least by a factor of
${M^2}/{m_W^2}$, where $M$ and $m_w$ are the masses of the $B_c$ meson
and the charged boson $W^\pm$ respectively. The precise contributions of the
SM, MHDM and RPV-SUSY models to the transition amplitude are given
in the following subsections.

\subsection{Standard model contributions}

The amplitudes arising from the Feynman diagrams figures 1-a,b,c are
given as

\begin{eqnarray}
{\cal A}_{a+b} = - \frac{i e G_F}{2 \sqrt 2} V_{cb}~ && \bar c
\biggr[Q_c \not\!{\epsilon} \frac{ \not\!{k} -\not\!{p_c} +m_c}
{(p_c \cdot k)} \gamma_\mu (1-\gamma_5) \nonumber\\
&&+ Q_b (1+\gamma_5) \gamma_\mu
\frac{\not\!{p_b} -\not\!{k} + m_b}{p_b \cdot k} \not\!{\epsilon}
\biggr] b\;
\bar \tau \gamma^\mu(1-\gamma_5)\nu_\tau \;,\label{eq:eqn1}
\end{eqnarray}

\begin{equation}
{\cal A}_c = -\frac{i e G_F}{2\sqrt 2} V_{cb}~ \bar c \gamma^\mu
(1-\gamma_5) b \; \bar \tau \biggr[\not\!{\epsilon} \frac{ \not\!{k} +
\not\!{p_l} +m_l}{(p_l \cdot k)} \gamma_\mu (1-\gamma_5)\biggr]
\nu_\tau \;\label{eq:eqn2}
\end{equation}
where, respectively, $Q_c$ and $p_c$ ( $Q_b$ and $p_b$) are the charge (in units of $e$)
and momentum of the constituent $c$ ($b$) quark, $\epsilon$ and $k$
denote the polarization and momentum of the photon and $p_l$ is the momentum
of the $\tau $ lepton.

Now we use the non-relativistic quark model to convert these quark level
amplitudes into the hadronic level. This model was previously used by us \cite{ref6} to
study the process $B_c \to l \bar \nu \gamma $. Since in the $B_c$ meson both
quarks ($b$ and $c$) are heavy, the relative momentum and their binding energy
to their masses are small, so the quark and anti-quark inside the
meson may be treated as approximately moving with small velocity. This means that
the following equations are valid to quite good accuracy:
\begin{equation}
M \simeq m_c + m_b\;,~~~~~ p_c \simeq \frac{m_c}{M} P~~~~{\rm and}~~~~
p_b \simeq \frac{m_b}{M} P\;
\end{equation}
where $P$ is the momentum of the $B_c$ meson. Using these
approximations and the interpolating field technique \cite{ref10}
to relate the hadronic matrix element to the decay constant of the
meson as

\begin{equation}
\langle 0| \bar c \gamma^\mu \gamma_5 b| B_c (P)\rangle =
i f_{B_c} P^\mu \;\label{eq:eqn3}
\end{equation}
we obtain the total amplitude for the $B_c \to \tau \bar \nu_\tau
\gamma $ process in the SM from equations (\ref{eq:eqn1}) and
(\ref{eq:eqn2}) to be:

\begin{eqnarray}
{\cal A}^{SM}& =& \frac{e G_F}{\sqrt 2} V_{cb} f_{B_c}
\biggr[m_{\tau} \bar \tau \biggr\{\frac{P \cdot \epsilon}{P \cdot k} -
\frac{\not\!{\epsilon} \not\!{k} + 2 p_l \cdot \epsilon}{ 2 p_l
\cdot k} \biggr\} (1-\gamma_5) \nu_\tau\nonumber\\
& +& \frac{1}{ 6 P \cdot k} \biggr\{ \left (\frac{M}{m_b}
-\frac{2 M}{m_c}\right )i \epsilon_{\mu\nu\alpha\beta} P^\nu k^\alpha
\epsilon^\beta\nonumber\\
&+& \left (6-\frac{M}{m_b} -\frac{2M}{m_c}\right )
\biggr( (P\cdot k) \epsilon ^\mu - (P\cdot \epsilon) k^\mu \biggr)
\biggr\} \bar \tau \gamma^\mu (1-\gamma_5) \nu_\tau \biggr]\;.\label{eq:eqn4}
\end{eqnarray}
It should be noted that, to ensure gauge invariance of the
transition amplitude exactly, we need to add a contact term which
causes the diagram figure 1-d with the propagators of the charged
$W$-boson to shrink into a `point'. Figure 1-d has one more
propagator of the charged gauge boson $W$, thus the gauge
violation, which originates from the ignoring of figure 1-d as stated above,
is suppressed by a factor of
$m^2_{B_c}/m^2_W$\cite{ref11}.

For convenience, we call the first term in equation (\ref{eq:eqn4}) as internal
bremsstrahlung part ($A_{IB}$), whereas the second term as the
structure dependent part ($A_{SD}$).

\subsection{Additional contributions from the multi Higgs doublet model}

In MHDM, the effective scalar-pseudoscalar
four-Fermi interaction can be induced by exchanges of additional
charged scalar Higgs particles with $CP$ violating complex couplings at the tree
level. To be more specific, here we consider the general MHDM\cite{ref12} with
natural flavour conservation.
In order to have observable $CP$ violation, the lightest charged
scalar particle has to be much lighter than the heavier ones\cite{ref13}.
Here we assume that all the charged scalar Higgs particles, apart from the lightest
one, decouple from the fermions effectively due to suppression
from their heavy masses in propagators. The Yukawa couplings of
the lightest charged scalar to up-type quarks, down-type quarks
and charged leptons are determined by the parameters $X$, $Y$ and
$Z$, respectively.

The fresh terms of the effective Lagrangian corresponding to the $B_c \to
\tau \bar \nu_\tau$ decay is

\begin{equation}
{\cal L} = \frac{G_F}{\sqrt 2} \frac{V_{cb}}{m_H^2}
\biggr\{ \bar c \biggr[m_b X (1+\gamma_5) +m_c Y (1-\gamma_5)
\biggr]b \biggr\}\biggr\{m_\tau
\bar \tau Z^* (1-\gamma_5) \nu_\tau\biggr\}\;\label{eq:eqn5}
\end{equation}
where $m_H$ denotes the mass of the lightest charged scalar particle.
In the above equation the term proportional to $m_c Y$ can be safely
neglected: first, it is suppressed by the mass ratio ${m_c}/{m_b}$;
second, $Y$ is bounded to be of ${\cal O} (1)$, while $X$ can be large
\cite{ref12}.

Now attaching the photon to the charged fermion lines of figure 2-a,b,c,
as in the SM case we obtain the quark level amplitudes to be

\begin{eqnarray}
{\cal A}_a^\prime+ {\cal A}_b^\prime = \frac{i e G_F R}{2\sqrt 2}
V_{cb}~ \bar c&&\biggr[ Q_c \not\!{\epsilon}
\frac{ \not\!{k} -\not\!{p_c} +m_c}
{p_c \cdot k} (1+\gamma_5) \nonumber\\
&&+ Q_b (1+\gamma_5)
\frac{\not\!{p_b} -\not\!{k} + m_b}{p_b \cdot k} \not\!{\epsilon}
\biggr] b\;\bar \tau (1-\gamma_5)\nu_\tau \;,\label{eq:eqn6}
\end{eqnarray}

\begin{equation}
{\cal A}_c^\prime = \frac{i e G_FR}{2\sqrt 2} V_{cb}~ \bar c
(1+\gamma_5) b \; \bar \tau \biggr[\not\!{\epsilon} \frac{ \not\!{k} +
\not\!{p_l} +m_l}{p_l \cdot k} (1-\gamma_5)\biggr]
\nu_\tau \;\label{eq:eqn7}
\end{equation}
where $R= X Z^*{m_b m_\tau}/{m_H^2}$.

These amplitudes can be converted to the hadronic level amplitudes
using the non-relativistic approximation (\ref{eq:eqn3}) and the
equation

\begin{equation}
\langle 0| \bar c \gamma_5 b| B_c(P)\rangle =
\frac{if_B M^2}{m_b+m_c} = i f_{B_c} M\;.\label{eq:eqn8}
\end{equation}
Again, by adding a contact term, we obtain the total gauge
invariant amplitude in the MHDM as

\begin{eqnarray}
{\cal A}^{MHDM} &=& {\cal A}_a^\prime+{\cal A}_b^\prime + {\cal
A}_c^\prime \nonumber\\ &=& \frac{e G_F R}{\sqrt 2} V_{cb}~
f_{B_c} M \bar \tau \biggr[ \frac{P\cdot \epsilon}{P\cdot k} -
(\frac{\not\!{\epsilon} \not\!{k} +2 (p_l\cdot \epsilon)}{2
p_l\cdot k})\biggr] (1-\gamma_5) \nu_\tau\;. \label{eq:eqn9}
\end{eqnarray}

It should be noted that the MHDM contributions only come from the
internal bremsstrahlung part (IB) and, as in
the case of SM, the diagram figure 2-d is necessary
to ensure exact gauge invariance, but with one more propagator
of the charged Higgs particle it is suppressed
by $m^2_{B_c}/m^2_{H^+}$.

\subsection{Additional contribution from the R-parity violating Model}

In the supersymmetric models there may be interactions which violate the baryon
number $B$ and the lepton number $L$ generically. To prevent the presence of
these $B$ and $L$ violating interactions in supersymmetric models, additional
global symmetry is required. This leads to the consideration of so-called
$R$-parity which is given by the relation $R_p =(-1)^{(3B+L+2S)}$,
where $S$ is the intrinsic spin of a field. Thus the $R$-parity can be used to
distinguish the particle ($R_p$=+1) from its superpartner ($R_p=-1$). Even
though the requirement of $R_p$ conservation gives a theory consistent with
present experimental investigations, there is no good theoretical justification for this
requirement; in particular, there is not very strong constraint whether
the lepton number $L$ is conserved or not. Therefore, models with explicit
$R_p$ violation ($\not\!{R_p}$) are considered by many authors\cite{ref14}.

The most general lepton number, so $R$-parity, violating super-potential is given by

\begin{equation}
W_{\not\!{L}} =\frac{1}{2} \lambda_{ijk} L_i L_j E_k^c
+\lambda_{ijk}^\prime L_i Q_j D_k^c \;\label{eq:eqn10}
\end{equation}
where, respectively, $i, j, k$ are generation indices, $L_i$ and $Q_j$ are
$SU(2)$ doublet lepton and quark superfields and $E_k^c$, $D_k^c$
are lepton and down type quark singlet superfields. Furthermore,
$\lambda_{ijk}$ is antisymmetric under interchange of the
first two generation indices.

The relevant Lagrangian for the decay mode $B_c \to \tau \bar \nu_\tau$
is

\begin{equation}
{\cal L}_{\not\!{R}} = - \frac{1}{2}\frac{\lambda_{3i3}\lambda_{i23}^
{\prime *}}
{ M^2_{\tilde {e}_{L_i}}}~ \bar c (1+\gamma_5)b ~\bar \tau
(1-\gamma_5) \nu\;\label{eq:eqn11}
\end{equation}
where the summation over $i=1,2$ is implied.

It should be noted that the $RPV$ Lagrangian has the same form as
the MHDM Lagrangian except for the couplings. From
equation (\ref{eq:eqn9}) we can thus easily obtain the amplitude for the
$B_c \to \tau \bar \nu \gamma$ in the $RPV$ model by replacing the
MHDM coupling ($ (G_F/\sqrt 2)V_{cb}R)$ by the RPV couplings
($-\lambda_{3i3}\lambda_{i23}^ {\prime *}/2 M^2_{\tilde {e}_{L_i}}
$), as shown in figure 3a-c\footnote{In fact, to be typical we consider one
slepton $\tilde{e}^+$ only for estimating the order.}.  We
will henceforth concentrate on the MHDM only and apply
the results to the RPV model with the above replacement.

It should be noted that, quite similar to the case of SM and MHDM, to
ensure gauge invariance of the transition amplitude exactly, figure 3-d
is necessary but with one more propagator of the slepton
$\tilde{e}^+$ it is suppressed by
$m^2_{B_c}/m^2_{\tilde{e}^+}$.

\section{Formalism for transverse $\tau$ polarization}

Having derived the transition amplitude within and beyond the
standard model, we now proceed to analyze the transverse polarization
of the $\tau $ lepton. Since within SM alone there is no observable
$CP$-violation effect at all, so we take into account the additional
contribution from the MHDH(RPV) model i.e. consider the amplitude SM+
MHDM(RPV) for the $B_c \to \tau \bar \nu_\tau\gamma$ decay
\begin{eqnarray}
{\cal A}^{SM+MHDM(RPV)} &=& \frac{e G_F}{\sqrt 2} V_{cb} f_{B_c}
\biggr[m_\tau\left (1 + R \frac{M}{m_\tau} \right )~
\bar \tau \biggr\{
\frac{P\cdot \epsilon}{P\cdot k} -
\frac{\not\!{\epsilon}\not\!{k} + 2(p_l \cdot \epsilon)}
{2 p_l \cdot k} \biggr\} (1-\gamma_5) \nu_\tau\nonumber\\
&+& \frac{1}{ 6 P \cdot k}\biggr\{ \left (\frac{M}{m_b}
-\frac{2 M}{m_c}\right )i \epsilon_{\mu\nu\alpha\beta} P^\nu k^\alpha
\epsilon^\beta\nonumber\\
&+& \left (6-\frac{M}{m_b} -\frac{2M}{m_c}\right )
\biggr( (P\cdot k) \epsilon ^\mu - (P\cdot \epsilon) k^\mu \biggr)
\biggr\} \bar \tau \gamma_\mu (1-\gamma_5) \nu_\tau
\biggr] \;.\label{eq:eqn13}
\end{eqnarray}

From this amplitude, a partial decay width and the transverse
polarization of $\tau$ are calculated.

To describe the decay $B_c \to \tau \bar {\nu}_\tau \gamma$ we
need two variables
\begin{equation}
x= \frac{2 P\cdot k}{ M^2}  \;\;\;\;  y =\frac{2 P \cdot p_l}
{M^2}\;.\label{eq:eqn14}
\end{equation}
In the centre mass frame of $B_c$, the variable $x(y)$ is
proportional to the photon (lepton) energy
\begin{equation}
x= \frac{2 E_k}{ M}  \;\;\; y =\frac{2 E_l}{M}\;.\label{eq:eqn15}
\end{equation}
The physical regions for $x$ and $y$ are given as follows
\begin{eqnarray}
&& 0 \le x \le 1-r \;\nonumber\\
&& 1-x + \frac{r}{1+x} \le y \le 1+r \;\label{eq:eqn16}
\end{eqnarray}
where $r={m_\tau^2}/{M^2}$. The Dalitz plot density is given by
\begin{equation}
\rho (x , y) \equiv \frac{ d^2 \Gamma}{dx dy} = \frac{M}{256
\pi^3} \sum_{spins} |{\cal A}|^2\;.\label{eq:eqn17}
\end{equation}
Here $\rho (x,y)$ has the form

\begin{equation}
\rho (x,y)= \rho_{IB} (x,y) + \rho_{SD} (x,y) +
\rho_{INT} (x,y)\;\label{eq:eqn18}
\end{equation}
\begin{eqnarray}
&&\rho_{IB} (x,y) = |1+ R \frac{M}{m_\tau}|^2 A_{IB}
f_{IB}\nonumber\\
&&\rho_{SD} (x,y) =\frac{2}{9M^2 x^2}  A_{SD} f_{B_c}^2
\biggr[\biggr(3-2\frac{M}{m_c}\biggr)^2 f_{SD}^+
+\biggr (\frac{M}{m_b}-3\biggr)^2
f_{SD}^-\biggr]\nonumber\\
&&\rho_{INT} (x,y) =\frac{2 }{3M x}A_{INT} f_{B_c}
\biggr(1+\frac{M}{m_\tau}
{\rm Re~(R)}\biggr)\biggr[\left (3-\frac{2M}{m_c}\right ) f_{INT}^+
+\left (\frac{M}{m_b}-3\right ) f_{INT}^-\biggr]\;\label{eq:eqn21}
\end{eqnarray}
where

\begin{eqnarray}
A_{SD} &=& \frac{M^5}{32 \pi^2} \alpha G_F^2 |V_{cb}|^2 \;\nonumber\\
A_{IB} &=& 2r \left (\frac{f_{B_c}}{M}\right )^2 A_{SD}\;\nonumber\\
A_{INT} &=& 2 r \left (\frac{f_{B_c}}{M}\right ) A_{SD} \;\label{eq:eqn22}
\end{eqnarray}
and

\begin{eqnarray}
&&f_{IB} (x,y)= \left (\frac{1-y+r}{x^2 (x+y -1 -r)}\right )
\biggr[x^2 + 2(1-x)(1-r)
-\frac{2xr(1-r)}{x+y-1-r}\biggr]\nonumber\\
&&f_{SD}^+ (x,y) = (x+y-1-r) \biggr[(x+y-1)(1-x)-r\biggr]
\nonumber\\
&&f_{SD}^- (x,y) =(1-y+r) \biggr[(1-x)(1-y) +r\biggr]\nonumber\\
&&f_{INT}^+ (x,y) = \left (\frac{ 1-y +r}{x(x+y -1-r)}\right )
\biggr[(1-x)(1-x-y)+r\biggr]\nonumber\\
&&f_{INT}^- (x,y) = \left (\frac{1-y+r}{x(x+y-1-r)}\right )
\biggr[x^2 - (1-x) (1-x-y)-r\biggr]\;.
\end{eqnarray}

If the transverse polarization $P_{T}$ of the $\tau$-lepton in the
decay $B_c \to \tau \bar \nu_\tau \gamma$ is defined
along the normal direction of the plane of the vectors $\vec k$
and $\vec {p}_\tau$, then

$$P_T\propto \vec {s_\tau}\cdot(\vec k \times \vec {p_\tau})$$
which is a $T$-odd or $CP$-odd quantity. If we introduce the unit
vector $\vec {n}_T =\frac {\vec k \times \vec {p_\tau}}{|\vec k
\times \vec {p_\tau} |}$, the transverse polarization of the
$\tau$-lepton at a fixed point in the phase space for the final
state is given as \cite{ref16}

\begin{equation}
P_T =\frac{d\Gamma (\vec {n}_T ) - d \Gamma (-\vec {n}_T )}
{d\Gamma (\vec {n}_T ) + d \Gamma (-\vec {n}_T )}\;\label{eq:eqn28}
\end{equation}
where $\vec k$ and $\vec {p}_\tau$ are the photon and $\tau$
momenta in the $B_c$ rest frame respectively, and $d\Gamma (\pm \vec {n}_T)$ are
the partial decay widths with the $\tau$ polarization along $\pm \vec
{n}_T$. Thus with the above equations (13-22) it is easy to find
\begin{equation}
P_T=-\sigma(x,y) {\rm Im}(R)
\end{equation}
with
\begin{eqnarray}
\sigma(x,y) &=& \frac{4 f_{B_c}^2}{3 M^2x} {A_{SD}}
\frac{\sqrt {(1-y+r)[(1-x)(x+y-1)-r]}}{\rho (x,y)}\;\nonumber\\
&\times & \biggr[\biggr(3-\frac{2M}{m_c}\biggr)
\biggr(\frac{(1-x)(x+y-1)-r}{x(x+y-1-r)}\biggr)
+\biggr(\frac{M}{m_b} -3\biggr)
\biggr(\frac{1+r-y}{x(x+y-1-r)}\biggr)\biggr]\;.\label{eq:eqn29}
\end{eqnarray}
The function $\sigma(x,y)$ is usually referred to as the distribution
of the transverse polarization of the charged lepton \cite{ref16}.
However, $P_T$ as defined above is not a direct observable. What
can be measured in a realistic experiment, for instance, is the
average of the polarization over the possible Dalitz plot regions,
which is given by
\begin{equation}
\bar P_T=\frac{\int_S dx~dy~ \rho(x,y) P_T(x,y)}
{\int_S dx~ dy~\rho(x,y)}\;.
\end{equation}
The average is a measure of the difference between the number of
$\tau$-leptons with their spins pointing above and below the decay
plane divided by the total number of $\tau$-leptons in the same
region of phase space. As final result, the averaged polarization is
obtained
\begin{equation}
\bar P_T=-\bar \sigma(x,y)~ {\rm Im}(R)\;.\label{eq:eqn00}
\end{equation}

\section{Numerical Results and Discussion}

Here we present the numerical analysis for the branching
ratio and transverse $\tau$ polarization in the $B_c \to \tau \bar
{\nu}_\tau \gamma $ decay process with the following values:
$m_c$ = 1.5 GeV, $m_b$ = 4.9
GeV, $M$ = 6.4 GeV, $m_\tau $ = 1.777 GeV, $f_{B_c}$ = 0.360 GeV,
$\tau_{B_c}$ = 0.46 ps, $|V_{cb}|$ =0.04 and $\alpha $ = 1/132.

We first estimate the branching ratio using equations (17 - 21). However, due to the
soft photon emission, the total decay width is singular when the photon energy
approaches to zero.  We therefore impose a cut value for the photon energy, which
will set an experimental limit on the minimal detectable photon energy. Since the
photon energy $E_k \ge $ 100 MeV, which corresponds to $x|_{min}$ = 3.125 $\times
10^{-2}$, we find the branching ratio in the standard model (R=0) to be
\begin{equation}
Br (B_c \to \tau \bar {\nu}_\tau \gamma) = 3.44 \times 10^{-4}\;.
\label{eq:eq1}
\end{equation}
This branching ratio may be accessible experimentally. Here we
note that the light front model framework yields \cite{ref17}
\begin{equation}
Br (B_c \to \tau \bar {\nu}_\tau \gamma) = 1.1 \times 10^{-4}\;.
\label{eq:eq2}
\end{equation}
If we compare equations (\ref{eq:eq1}) and (\ref{eq:eq2}), it is clear that the light
cone prediction on the branching ratio is approximately three times smaller than
our predicted value. Although right now no one can tell the precise reason what
makes the difference, one may see from the definition of $P_T$ Eq. (\ref{eq:eqn28})
(being a fraction) that the difference should be canceled quite a lot in $P_T$.

For MHDM, we use the following parameters
from \cite{ref12}. The perturbative constrains on $|XZ|$ and
$Im(XZ^*)$ are obtained from the decay $B \to X \tau \nu_\tau $.
Namely in MHDM, the branching ratio for this decay mode is given as
\cite{refa1}
\begin{equation}
BR^{MHDM}(B \to X \tau \bar \nu_{\tau})=BR^{SM}
(B \to X \tau \bar \nu_{\tau})\left (1+\frac{|R|^2}{4}
-D \mbox{Re}(R) \right )\;,\label{eq:m5}
\end{equation}
where the Standard Model result is \cite{refa2}
\begin{equation}
BR^{SM}(B \to X \tau \bar \nu_{\tau})=(2.30 \pm 0.25)\%
\end{equation}
and $R=m_\tau m_b X Z^*/m_H^2 $ defined as in Eqs. (\ref{eq:eqn6,eq:eqn7}).
The expression for $D$ is given in
\cite{ref12}. Comparing Eq. (\ref{eq:m5}) with the experimental
value\cite{refa3}
\begin{equation}
BR(B \to X \tau \bar \nu_{\tau})<4\%    ~~~~~95\% ~\mbox{CL.},
\end{equation}
one will obtain a perturbative constraint on $|XZ|$ when $m_{H^\pm} > 370$
GeV as follows
\begin{equation}
|XZ| <{\rm min} ( 0.32~ m_{H^\pm}^2~ {\rm GeV}^{-2} , 44200)\;.
\label{eq:eqn001}
\end{equation}
Thus the accordingly up-bound on $Re(XZ^*)$ is the same as that on $|XZ|$.

However the CP violating parameter $\mbox{Im} (XZ^*)$ can be bounded
from the transverse polarization of the muon in the decays
$K^+ \to \pi^0 \mu^+ \nu_\mu $ and $B \to X \tau \bar \nu_{\tau}$.
When $m_{H^\pm} \geq 440 $ GeV the perturbative constraint on $Im(XZ^*)$
is given as
\begin{equation}
{\rm Im} (XZ^*) < {\rm min} (0.23~ m_{H^\pm}^2~ {\rm  GeV}^{-2}, 44200)\;.
\label{eq:eqn002}
\end{equation}

With these constraint values of Eq. (\ref{eq:eqn001}) and
Eq. (\ref{eq:eqn002}), we may evaluate the up-bounds for the
branching ratio and averaged transverse polarization of MHDM
precisely by integrating
Eq. (\ref{eq:eqn17}) with respect to $x$ and $y$ within the limits given by
Eq. (\ref{eq:eqn16}) and Eq. (\ref{eq:eqn00}). Finally, we obtain quite
interesting results and put them in Table I.

\begin{center}
{\bf Table I.} The dependence of the up-bounds for the averaged transverse polarization
of $\tau$ lepton $\bar{P_T}$ (in $10^{-2}$) and the branching ratio
$B_c\to \tau\nu_\tau\gamma$ (in $10^{-3}$) on the mass of the lightest charged Higgs
$m_H$ (in GeV) in MHDM with the constraint Eq. (\ref{eq:eqn001}) and Eq.(\ref{eq:eqn002})
obtained from the decays
$K^+ \to \pi^0 \mu^+ \nu_\mu $ and $B \to X \tau \bar \nu_{\tau}$.
\vspace{4mm}

\begin{tabular}{c|c|c|c|c|c}
\hline\hline
$m_H$ in GeV & Br($B_c\to \tau\nu_\tau\gamma$)& $\bar{P_T}$ &
$m_H$ in GeV & Br($B_c\to \tau\nu_\tau\gamma$)& $\bar{P_T}$ \\
in GeV & in $10^{-3}$ & in $10^{-2}$ &
in GeV & in $10^{-3}$ & in $10^{-2}$\\
\hline
370 & 9.05 & - & 400 & 6.81 & - \\
\hline
440 & 4.91 & 14.8 & 450 & 4.55 & 15.2 \\
\hline
460 & 4.23 & 15.6 & 470 & 3.94 & 16.1 \\
\hline
480 & 3.67 & 16.5 & 500 & 3.22 & 17.4 \\
\hline
600 & 1.86 & 20.8 & 700 & 1.245 & 22.9 \\
\hline
800 & 0.945 & 23.6 & 900 & 0.743 & 23.2 \\
\hline
1000 & 0.632 & 22.1 & 1100 & 0.557 & 20.7 \\
\hline
\hline
\end{tabular}

\end {center}

We may see from Table I. that the up-bound of the branching ratio for the radiative
leptonic decay

\begin{equation}
Br (B_c \to \tau \bar {\nu}_\tau \gamma) < 9.05 \times 10^{-3}
\end{equation}
which is quite big\footnote{Here
the constraint on the parameters $X, Z$ quoted from Ref.\cite{ref12} in MHDM is very weak
and the contributions from the lightest charged Higgs particle, being comparatively light,
are additive to SM, so a quite big up-bound value of the branching is obtained.}.
Similarly from the table the up-bound of the averaged transverse polarization of $\tau$
lepton obtained from Eq. (\ref{eq:eqn00}) can be

\begin{equation}
\bar P_T < 15\sim 23 \%
\end{equation}
The $\tau$ polarization was estimated by Geng et al\cite{ref17} in MHDM and
obtained a up-bound $P_T < 0.67$. Since the effect of MHDM in $A_0 (x,y)$
($P_T=A_T(x,y)/A_0(x,y)$ in their definition) was not taken into account,
so such a big up-bound was obtained.

As for RPV, when assuming the sfermion masses to be 100 GeV the R-parity
violating couplings are obtained by Ref.\cite{ref18}
\begin{eqnarray}
&&\lambda_{313} = -0.003 \;\;\;\; \lambda_{323} = -0.06 \nonumber\\
&&\lambda^\prime_{123} =0.20 \;\;\;\;\;~~ \lambda^\prime_{223}=0.18 \;. \label{eq:eqn003}
\end{eqnarray}

Assuming maximal $CP$ violation further and with the couplings Eq. (\ref{eq:eqn003}), we
may evaluate the up-bounds on the branching ratio and $\bar P_T$ of RPV precisely,
and put the results in Table II.

\begin{center}

{\bf Table II.} The dependence of the up-bounds for the averaged transverse polarization
of $\tau$ lepton $\bar{P_T}$ (in $10^{-4}$) and the branching ratio
$B_c\to \tau\nu_\tau\gamma$ (in $10^{-4}$) on the mass of the lightest charged slepton
$m_{\tilde{l}}$ (in GeV) in RPV with the couplings Eq. (\ref{eq:eqn003}).

\vspace{4mm}
\begin{tabular}{c|c|c|c|c|c}
\hline\hline
$ m_{\tilde{l}}$ & Br($B_c\to \tau\nu_\tau\gamma$)& $\bar{P_T}$ &
$ m_{\tilde{l}}$ & Br($B_c\to \tau\nu_\tau\gamma$)& $\bar{P_T}$ \\
in GeV & in $10^{-4}$ & in $10^{-4}$ &
in GeV & in $10^{-4}$ & in $10^{-4}$\\
\hline
100 & 3.451 & 9.77 & 200 & 3.447 & 2.59 \\
\hline
300 & 3.447 & 1.15 & 400 & 3.446 & 0.65 \\
\hline
500 & 3.446 & 0.04 &  &  &  \\
\hline
\hline
\end{tabular}
\end {center}

In summary, for RPV the up-bound of the branching ratio is
\begin{equation}
Br (B_c \to \tau \bar {\nu}_\tau \gamma) < 3.45 \times 10^{-4}
\end{equation}
and the up-bound of the averaged transverse polarization is
\begin{equation}
\bar P_T > 0.1\%\;.
\end{equation}

The great differences of MHDM from RPV in the up-bound values for the branching ratio
and the averaged transverse polarization are due to the facts that i). The constraint
for MHDM Eq. (\ref{eq:eqn001}) and Eq. (\ref{eq:eqn002}) is very weak, but the constraint
on the couplings
Eq. (\ref{eq:eqn003}) for RPV is quite strained and small. ii). The additional contributions
from the charged Higgs in MHDM are constructive (plus) to those of SM, but the additional
ones from the charged slepton are destructive (minus) to those of SM.

It is interesting to compare the sensitivity of transverse $\tau$
polarization in $B_c \to \tau \bar {\nu}_\tau \gamma$ to that of
the muon in $K^+ \to \mu \bar {\nu}_\mu \gamma$.  {\it A priori}
we expect the polarization effects to be larger for $B_c \to
\tau \bar {\nu}_\tau \gamma$ than  $K^+ \to \mu \bar {\nu}_\mu
\gamma$ since there are larger quark and lepton masses in the $B_c$ case.
For example, the lepton polarization in these two cases may
generically be written in MHDM as $\bar P_T \sim -\sigma_l~{\rm
Im} (R)$. The ratio $R^\tau/R^\mu$ is enhanced roughly by the
factor $M m_\tau / m_K m_\mu$, thus the transverse lepton
polarization is enhanced by $P_\tau/P_\mu \sim M m_\tau/ m_K m_\mu
\sim 2 \times 10^2$. The rather large enhancement which we
generically find means that, to reach a given
sensitivity for new physics, we require far fewer $B_c$ decays
than $K$ decays, although it is a
great challenge for experiment to produce such a great amount of $B_c$
mesons and reject possible background.

In conclusion, we have estimated the branching ratio and $T$-odd
transverse polarization of $\tau$ leptons in the $B_c \to \tau \bar
{\nu}_\tau \gamma$ decay in the MHDM and RPV models. It is encouraging
to note that the polarization effects in these models are found to
be quite large. The transverse $\tau$ polarization observable may provide a means
to look for the effects of new physics, since it has received a
negligible contribution from the standard model sources. Based on the above
estimate for the averaged $\bar P_T$ and branching ratio, we can conclude that
at the level, $n$ times $\sigma$ (the standard deviation), and with 4\% efficiency
in detecting the $\tau$ lepton, the observations of the branching ratio and the
averaged $\tau$ transverse polarization experimentally may make substantial
improvements on present constraint obtained by the best observations on suitable
observables for MHDM and RPV, if the numbers of $B_c$ mesons so huge as
$4.3\cdot n^2 \times 10^{5}$ for MHDM and $7.2 \cdot n^2 \times 10^{10}$
for the RPV model are collected. Namely, at LHC one must obtain meaningful results
for MHDM and RPV, even at Tevatron with RUN-II data one may be possible to obtain
some meaningful result for MHDM if one observes the branching ratio and transverse
polarization very carefully.

\acknowledgements We would like to thank the organizers of
WHEPP-6, held at the Institute of Mathematical Sciences, Chennai,
India, for providing a stimulating atmosphere, where $B$ physics
was discussed in a working group. A.K. Giri would like in particular to thank
C.S.I.R., Government of India for financial support. C.-H. Chang and G.-L Wang
would like to thank National Natural Science Foundation of China (NSFC) because
this work was supported in part by NSFC.

\begin{figure}
\begin{center}
   \epsfig{file=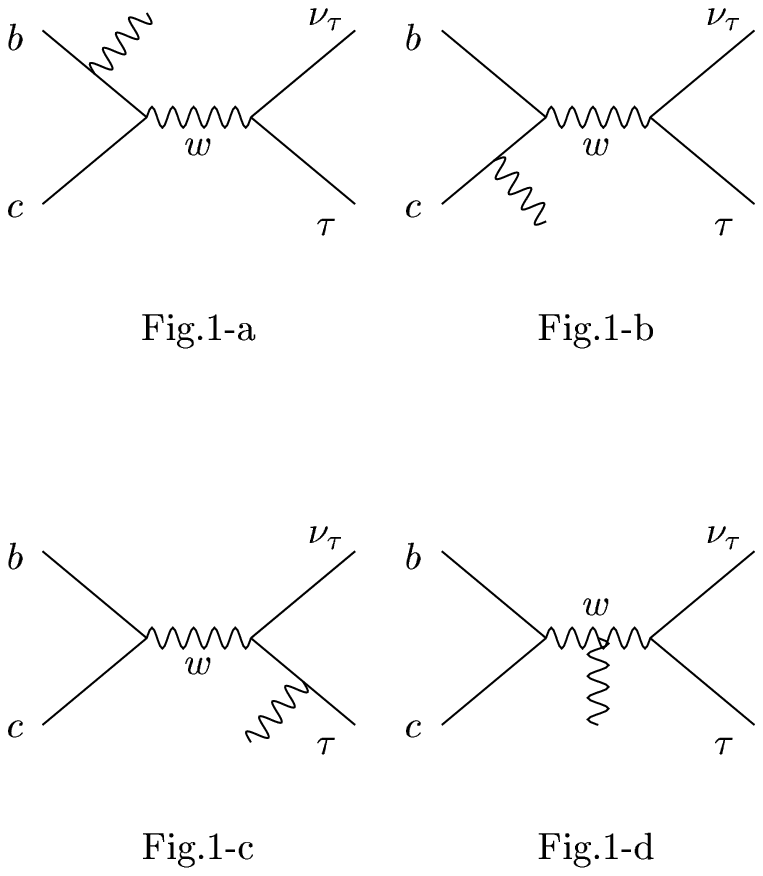, bbllx=100pt,bblly=260pt,bburx=450pt,bbury=570pt,
width=10cm,angle=0} \caption{Feynman diagram for the weak
boson exchange.}
\end{center}
\end{figure}

\begin{figure}\begin{center}
   \epsfig{file=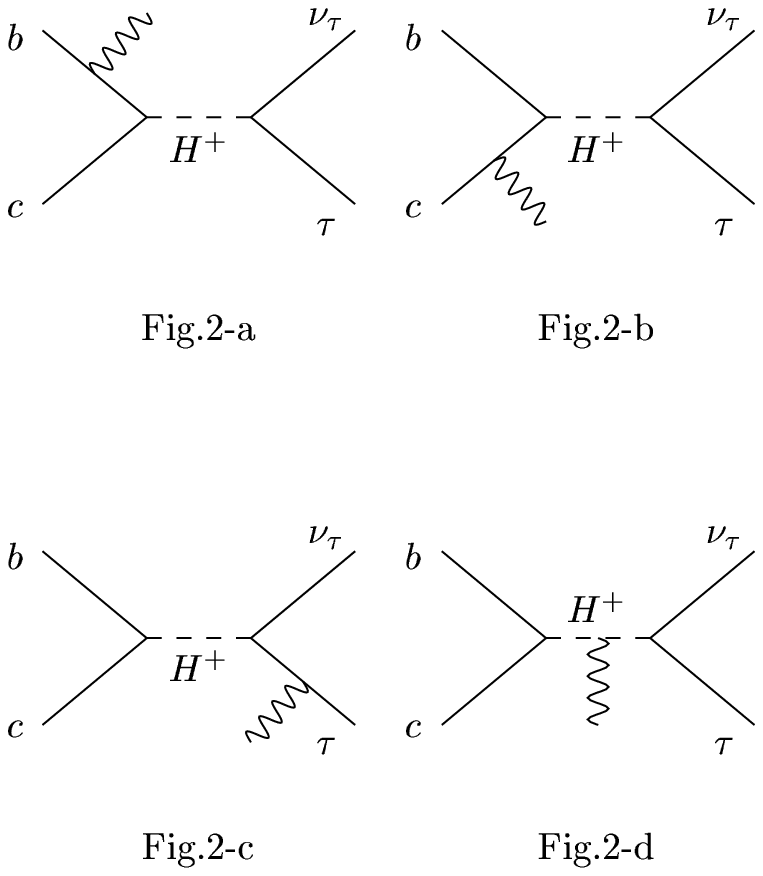, bbllx=100pt,bblly=260pt,bburx=450pt,bbury=570pt,
width=10cm,angle=0} \caption{Feynman diagram for the lightest
charged Higgs exchange.}
\end{center}
\end{figure}

\begin{figure}\begin{center}
   \epsfig{file=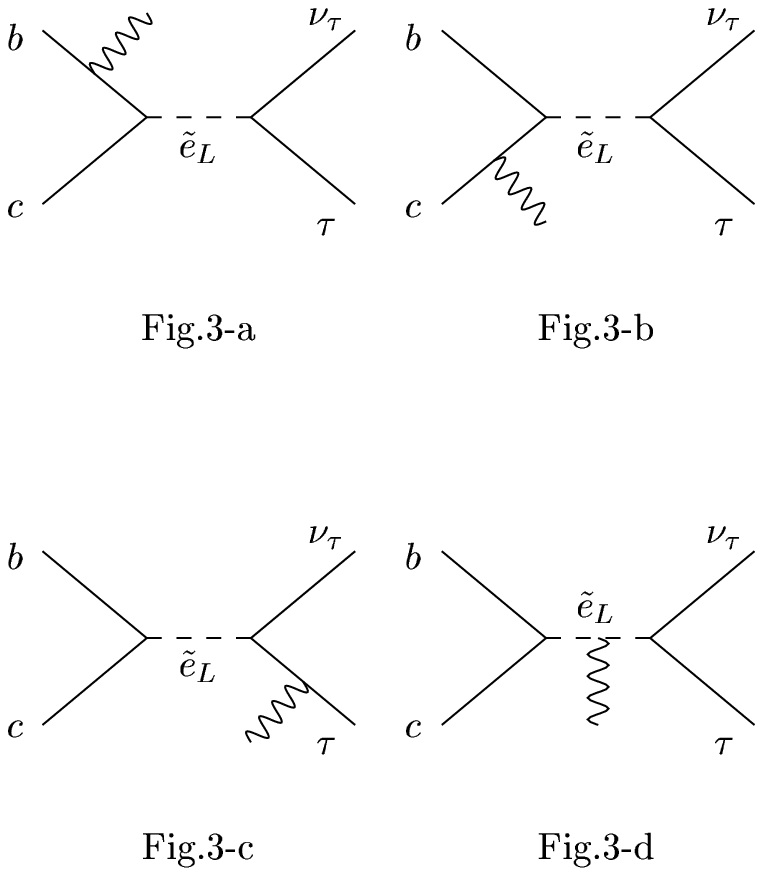, bbllx=100pt,bblly=260pt,bburx=450pt,bbury=570pt,
width=10cm,angle=0} \caption{Feynman diagram for the lightest
charged slepton exchange.}
\end{center}
\end{figure}

\end{document}